\documentclass[
prb,preprint%
,showpacs]{revtex4}
\usepackage{graphicx}
\usepackage{dcolumn}
\usepackage{bm}
\begin{document}
\title{Effect of CH$_{4}$ addition on excess electron mobility in
liquid Kr}

\author{A.F.Borghesani}\email{borghesani@padova.infm.it}
\affiliation{Istituto Nazionale per la Fisica della Materia \\
 Department of Physics, University of Padua\\
 Via F. Marzolo 8, I-35131 Padua, Italy}

 \author{}\affiliation{Istituto Nazionale di Fisica 
 Nucleare\\ Department of Physics, University of Bologna\\ via C. Berti Pichat, 
 6/2, I--40127 Bologna, Italy}
 
\author{M.Folegani\footnote{Present address: AIM S.p.A., v. Ripamonti 129,
I-20129 Milan, Italy}, P.L.Frabetti\footnote{Permanent address: Istituto Nazionale di Fisica 
 Nucleare, Department of Physics, University of Bologna, via C. Berti Pichat, 
 6/2, I--40127 Bologna, Italy}, and L.Piemontese}
\affiliation{Istituto Nazionale di Fisica Nucleare,\\ Department of 
Physics, University of Ferrara\\ via Paradiso 12, I--44100 Ferrara, Italy}%

\begin{abstract}
The  excess electrons mobility $\mu$ has been measured recently 
in liquid mixtures of Kr and 
CH$_{4}$ as a function of the electric field up to 
$E\approx 10^{4} \, V/cm $ and of 
the CH$_{4}$ concentration $x$ up to $x \approx 10 \% ,$ at 
temperatures $T\approx 130 \,K,$ 
fairly close to the normal boiling point of Kr 
$(T_{b}\approx 120\, K)$~\cite{folegani}. We present here new data 
which extend the previous set in the region of low electric field.
The experimental results are interpreted in 
terms of a kinetic model previously proposed to explain the concentration 
dependent behavior of $\mu$ in liquid Ar--Kr and Ar--Xe 
mixtures. The main result is that CH$_{4}$ is more effective in 
enhancing energy--transfer rather than  momentum--transfer in comparison 
with mixtures of liquified noble gases. 
The 
field dependence of $\mu$ is quite complicate. 
In particular, at intermediate values of the field, there appears to be a 
crossover between two different electric--field dependent behaviors of 
$\mu.$
The electric field strength at crossover is well correlated with the 
concentration of CH$_{4}.$ 
This fact suggests that different excitations of the molecular solute 
might be involved in the momentum-- and energy--transfer processes for 
different values of the mean electron energy.
\end{abstract}

\maketitle

 \section{Introduction}\label{sec:intro} 
The study of electron conduction in nonpolar liquids is of great 
fundamental and technical interest \cite{holsch89,hatano97}. Non polar 
liquids are examples of 
simple disordered materials. Theoretical investigations of the 
electronic states in such media are connected to experiment by the low 
field behavior of the excess electron mobility \cite{finn86}. The 
properties of excess electrons in such media may also give useful 
information on the electronic states in noncrystalline solids 
\cite{doke,yoshino76,sowada77}. 

Non polar liquids are also used as sensitive media in 
high--energy physics ionization detectors. 
For such an application it is essential 
to know the behavior of excess electrons in the high--field region,
where detectors are typically operated, in order to 
understand the 
basic electronic conduction mechanisms in that particular range 
for the goal of optimizing the detectors' performances \cite{schmidt97}.

Liquified heavy noble gases, Ar, Kr, and Xe, are mainly used in 
ionization chambers because of the combination of large density and high 
values of the excess electron mobilities. At small electric fields 
mobilities exceeding several hundreds of $cm^{2}/Vs$ have been 
measured~\cite{miller,schnyders,kf,halpern,freemanxe,eibl,lamp}.  
In these so--called high--mobility liquids electrons are assumed to 
be quasifree because their mean free path is relatively long although 
they are moving in a very dense environment of atoms which are very 
effective scattering centers in the gas phase \cite{holsch89}. 

At small electric fields electrons are in near thermal equilibrium 
with the atoms of the host medium. The electron mobility is determined by 
processes of elastic scattering and is approximately independent of 
the field strength. In this region the drift velocity increases 
linearly with the field strength. At higher values of the field the 
electron drift velocity increases less than proportionally with the 
field and, finally, it nearly saturates. This behavior is related to the 
increase of the mean electron energy with the field and to the 
dependence of the scattering rate on the electron energy.

It has been shown \cite{yoshino76,sakai,rahm,seguinot,vuillemin,BIC} that the 
addition of a molecular or atomic solute in small 
proportion influences the 
dependence of the mobility on the field strength, especially at high 
fields. In particular, the saturation drift velocity is largely 
increased above the value in the pure liquid. This effect is commonly 
explained as due to a reduction of the electron mean energy upon additional 
inelastic scattering with the impurities. These act as 
additional scattering centers that are very effective in thermalizing 
electrons even at quite large field strengths. In particular, for a 
spherical symmetric molecule like CH$_{4},$ the inelastic processes 
are assumed to be due to the excitation of the main vibrational levels 
of the molecule (0.16 and 1.75 $eV)$~\cite{sakai,sakai2} upon collisions with energetic enough electrons .

The search for the optimum combination of base liquid and impurity  
for the best detector performance is 
far from being concluded \cite{rahm,seguinot,vuillemin}. 

These mobility measurements are obviously important for detectors' 
designers. However, from a fundamental point of view these data may 
give additional hints for 
the description of the electron mobility in liquids, which is not yet 
completely understood~\cite{holsch89}. 
In fact, the theoretical treatment of the field behavior of the mobility in 
a pure liquid is far from being satisfactory. 

The description of the 
electron mobility in the liquid can be approached from two completely 
opposite directions. The first approach is the so called {\it single--scattering}
picture, essentially due to Lekner~\cite{lek1,lek2}, in which 
electrons scatter off single atoms of the liquid or, more precisely, 
they undergo a binary collision with a single interaction potential that 
takes into account contributions from the potentials of nearby atoms. 
In this model, the  
two--term solution of the Boltzmann transport equation 
\cite{hux,phelps} allows the calculation of an energy--dependent 
scattering cross section.

At the opposite extreme, the {\it 
deformation potential theory} 
of Basak and Cohen (BC)~\cite{BC} represents the motion of the 
quasi--free electrons in the liquid as that of a wave in a 
quasi--periodic structure.  In this model the electron mobility 
in a liquid is determined by the scattering produced by fluctuations of 
the bottom of the conduction band due to the intrinsic density 
fluctuations of the liquid.

In the latter model, the addition of a solute is assumed to enhance 
the disorder inherent to a fluid by introducing concentration 
fluctuations in addition to the usual density fluctuations. In the 
former one, many--body and multiple scattering effects are included in the 
cross sections, which have now to be determined by a fit to the 
experimental data. 

On one hand, the BC model has been adapted with reasonable success to 
mixtures of liquid hydrocarbons, 
where both solvent and solute are liquid at the same 
temperature~\cite{kengo}. Unfortunately, its predictions have proven 
completely wrong when it is used for the description of the electron 
mobility in liquid mixtures of noble gases, namely Kr or Xe in liquid 
Ar~\cite{BIC}. 

A further drawback of the BC model, and of its 
extension to mixtures, is that it makes predictions only on the 
zero--field mobility and does not describe its electric--field 
dependence, which, on the contrary, is very important 
from the point of view of detectors' design.
 
On the other hand, the single--scattering (or gas--kinetic) approach 
has been succesfully used to describe the electric field dependence 
of the electron mobility in liquid Ar 
and methane \cite{kaneko,kaneko2} by introducing two 
constant scattering cross sections, the momentum-- and 
energy--transfer cross sections, in the wake of Lekner's theory. 
In this model, the many--body and multiple 
scattering effects due to the combined effect of short interatomic 
distances and large electron wavelength are embodied in the effective 
cross sections. This dressing of the scattering 
cross section due to multiple scattering effects is also at the base of the heuristic kinetic model, 
developed for the successful description of the electron mobility 
in dense noble 
gases \cite{borg1,borg2,borg3,borg4}.

The same gas--kinetic approach has proven also quite succesful 
in the case of mixtures of liquified noble gases~\cite{BIC}. Although 
in pure Ar and CH$_{4}$ the choice of two constant cross sections 
provides a quite nice agreement with the experimental 
data~\cite{kaneko,kaneko2}, an even better description of the electron 
mobility as a function of the electric field for all impurity 
concentrations in the mixtures of liquified noble gases 
is obtained by assuming a constant momentum transfer cross section 
$\sigma_{m}$ and an energy--transfer cross 
section $\sigma_{E}$ inversely proportional to the electron energy 
$\epsilon, $
$\sigma_{E}\propto 1/\epsilon$ \cite{BIC}. This 
dependence has been chosen only on a phenomenological basis and its 
meaning is not clear yet. However, since in this gas--kinetic model 
the cross sections are treated as adjustable parameters, it can be 
only said that this energy--dependent energy--transfer cross section 
provides a much better fit to the experimental data than a constant 
cross section does.

In this work we 
have therefore more carefully investigated 
the effect of the addition of the molecular solute 
CH$_{4}$ to pure liquid Kr, with emphasis on the low-- and 
intermediate--field behavior of 
the mobility. Preliminary measurements, especially 
concerned with the mobility behavior at 
high fields, have been reported previously~\cite{folegani}, keeping in 
mind the application of such mixtures in ionization detectors. 
We report here a more complete study that includes new measurements at 
quite small electric field strengths aimed at a more physical goal.

This goal is twofold. On one hand, this 
mixture has never been studied before and might be promising as a medium 
for ionization detectors. On the other hand, it represents a different
benchmark for the validation of the gas--kinetic model and the 
determination of the concentration dependence of the effective cross 
sections might contribute useful pieces of information on the 
effectiveness of elastic and inelastic electron scattering processes 
in the liquid.

\section{Experiment}
The experimental apparatus has been described 
elsewhere \cite{folegani} and we refer to literature for the details. 
We recall here only the essential features of the experiment. The 
cell is a typical double--gridded ionization chamber. Excess electrons 
are photoextracted from a Ni--coated brass cathode by a 
short pulse of ultraviolet 
light produced by a Xe flashlamp. The four electrodes, anode, cathode 
and the two grids, are kept at suitable voltages in order to ensure 
the maximum grid transparency \cite{bunne}. Several guard rings are 
kept at the appropriate voltage by a resistor cascade in order to 
ensure the maximum field uniformity in the cylindrical drift space.  

The electrodes are connected to charge amplifiers. The signal 
induced by the drifting electrons is recorded by a digital oscilloscope and 
analyzed by means of a personal computer. The drift time can be easily 
determined by analyzing the signal shape. To span a large region of 
electric fields in a single sweep both the drift times between the 
cathode and the first grid and between the first and the second grid 
are 
recorded simultaneously. The drift time in the region between the 
second grid and the anode, 
owing to the short distance and to the strong field 
between these two electrodes, cannot be measured reliably and has not 
been recorded. 
The two sets of drift mobility measured in the first and second 
region agree well with each other within the experimental accuracy 
where the field values overlap. The overall accuracy of the mobility 
measurements is better than 10 \%.

Very pure, commercially available 
gaseous mixtures of CH$_{4}$ in Xe of different and known 
composition are condensed in the cell through an Oxisorb purifier to 
remove oxygen and water vapor impurities. 
The cell is cooled down to the desired temperature by immersion in an  
isopentane bath cooled by liquid N$_{2}.$ Two thermoresistors located 
in the cell near the cathode and the anode, respectively, provide 
constant monitoring of the temperature. The temperature is stable 
within $\pm 0.5 \, K.$

\section{Experimental Results and Discussion}
The excess electron mobility in liquid Kr has been measured as a function of the 
electric field strength up to $E\approx 10 \, kV/cm$ at $T\approx 130 \, 
K$ 
for 
several mixtures of different composition. The liquid density of pure 
Kr at the temperature of the experiment is $N\approx 167
\times 10^{26}\, m^{-3}.$ The CH$_{4}$ concentration, 
$x,$ of the mixtures is $x =$ 0.01, 0.1, 0.5, 1, 2, 3, 5, and 10 
\%, respectively. 
As a calibration we have also measured the mobility in pure liquid 
Krypton. Our data in the pure liquid agree well with literature data 
\cite{finn86}.

We report the mobility as a function of the 
electric field strength $E$ for several of the mixtures under 
investigation in Figure 1 and Figure 2 
in order to avoid overcrowding of the figures. 
All mixtures show similar features.

As usual, the electron mobility $\mu$ shows a low--field behavior 
where it is essentially independent of the electric field. In this 
region the excess electrons are in near thermal equilibrium with the 
atoms of the liquid, do not gain very much energy from the 
electric field, and  mainly undergo elastic collisions 
that determine the mobility.

At higher fields, the mobility depends on the field strength and 
decreases sharply with increasing field. In this region the  
mean electron energy is greatly enhanced by the field and the net effect is a 
large  
increase of the scattering rate, leading to the observed decrease of 
the mobility. 
This behavior is common to all mixtures with some important 
differences. 

At small concentration of methane the effect of the 
solute on the zero--field mobility $\mu_{0}$ is not very large. 
Only for the highest CH$_{4}$ concentration $(x\approx 10 \,\% )$, 
there is a significant reduction of $\mu_{0},$ as shown in 
figure 3.
\noindent Small concentrations, as evident from Figures 1 and 2, do mostly 
influence the high--field behavior of the mobility. An influence 
on momentum transfer is obtained only at high methane 
concentrations. This behavior is similar to that observed in mixtures 
of light alkanes in liquified noble gases \cite{yoshino76}. However, 
this is the first time for this behavior to be observed in the 
present mixture of CH$_{4}$ in liquid Kr.

Moreover, it has to be noted 
that the situation in mixtures of liquified noble gases is completely 
different as far as $\mu_{0}(x)$ is concerned \cite{BIC}. In fact, in the  
liquid Ar--Kr and liquid Ar--Xe mixtures, $\mu_{0}$ decreases rapidly 
with increasing solute concentration, especially in the Ar--Xe mixture. 
In the Ar--Kr mixture, where a concentration of $\approx 30 \,\%$ is 
reached, $\mu_{0}$ eventually levels off and becomes nearly 
concentration independent. In the Ar--Xe mixture, for concentration up 
$x\approx 
5\, \% ,$  $\mu_{0}$ decreases linearly with large neative slope as $x$ 
is increased.

The sharp decrease of $\mu$ with increasing $E$ at larger fields 
(see Figures 1 and 2) is commonly attributed 
to the increase of the mean electron energy with increasing electric 
field. Between collisions, at higher fields, electrons pick up more 
energy from the field than they are able to share with the 
liquid upon collisions and thus become epithermal. 
The scattering rate is consequently enhanced 
and the mobility decreases. 
 
The change of the electric field dependence of the mobility (see 
Figures 1 and 2) is therefore due to the increase of the mean 
electron energy above the thermal value because of the applied 
electric field. Upon increasing the methane concentration in the 
mixture, the region where the mobility is field independent spans 
a wider field range and the transition to the hot--electron behavior 
shifts to larger field strengths. 
 Moreover, in the epithermal region, higher field strengths are 
required, upon increasing the impurity concentration, 
in order to achieve the same mobility value. Since $\mu$ depends on 
the mean electron energy only, provided that all other parameters are 
kept fixed, this fact means that electrons reach the 
same mean energy at larger fields for increasingly higher concentration 
of impurities.

These observations confirm the 
assumption that molecular impurities are very effective in 
thermalizing electrons at larger fields although they are not as 
effective as atomic impurities as far as momentum transfer is concerned. 
Molecular impurities act as additional scattering 
centers for electrons where they might lose energy in inelastic 
collisions more efficiently than in the pure liquid. 

A similar behavior has been observed also in liquid mixtures of 
liquified noble gases \cite{BIC}, where atomic impurities Xe and Kr 
were dissolved in liquid Ar. Also in that case, atomic impurities 
extend the range of the thermal behavior of electrons. However, a 
stronger dependence of $\mu_{0}$ on the impurity concentration was 
observed. This behavior is probably related to the fact that atomic 
impurities have fewer inelastic scattering channels for low--energy 
electrons than molecular impurities do and affect more effectively 
the momentum--transfer--  rather than the energy--transfer processes. 

At high fields the drift mobility is determined essentially by the 
relative magnitudes of elastic and inelastic energy loss rate. When 
inelastic losses become greater than the elastic ones the drift 
mobility increases above the value of the pure liquid.

In order to carry out a simple analysis of the experimental data, we 
adopt the gas--kinetic approach of Kaneko {\it et al.}, based on the 
two--term solution of the Boltzmann transport 
equation \cite{kaneko,kaneko2}.
For a simple fluid of number density $N,$ the mobility $\mu$ 
is given by the usual formula 
\cite{lek1,lek2,hux}

\begin{equation}
    \mu = -\left( {e\over 3}\right) \left( {2\over m}\right) ^{1/2}
    \int\limits_{0}^{\infty}{\epsilon\over 
    N\sigma_{m}\left(\epsilon\right) }
    \left[{\mathrm{d}g\left(\epsilon\right) \over \mathrm{d}\epsilon}
    \right]\,\mathrm{d}\epsilon
    \label{eq:eq1}
\end{equation}
where $\sigma_{m}(\epsilon)$ is the energy--dependent 
momentum--transfer scattering cross section. $e$ and $m$ are the 
electron charge and mass, respectively. 
The Davydov--Pidduck electron energy distribution function 
$g(\epsilon)$ is given by
\begin{equation}
    g\left(\epsilon\right) = A \exp{\left\{-\int\limits_{0}^{\epsilon}
    {\mathrm{d}z\over k_{\mathrm{B}}T +\left( {M\over 6m}\right)
    \left( {eE\over N}\right)^{2}{1\over z \,\sigma_{m}\left( 
    z\right) \sigma_{E}\left(z\right) }
    }
 \right\}}
    \label{eq:eq2}
\end{equation}
where $M$ is the atomic (molecular) mass, 
$\sigma_{E}\left(\epsilon\right)$ is the energy--dependent 
energy--transfer scattering cross section. The constant $A$ is 
fixed by the normalization condition $\int_{0}^{\infty}\sqrt{z} 
g\left(z\right) \mathrm{d}z =1.$

From Eqns. \ref{eq:eq1} and \ref{eq:eq2} it is evident that the 
mobility is primarily determined by the momentum--transfer cross 
section, but the energy--transfer cross section also
affects the mobility because it influences the electron energy 
distribution function by controlling the rate at which energy is 
exchanged.

Strictly speaking, in order to account properly for the scattering of 
electrons off correlated atoms in the fluid \cite{lek3}, Eq. \ref{eq:eq1} 
should be divided by the long--wavelength limit $S(0)$ of the static 
structure factor, that takes into account the compressibility of the 
medium. However, according to Kaneko {\it et al.}, we set $S(0)=1$ and 
every correlation effect is now accounted for by the effective cross 
section determined in this way.

By introducing constant values for the two relevant cross section 
$\sigma_{m}$ and $\sigma_{E}\approx 100 \sigma_{m},$ this model 
reproduces quite well the low--field limit of the experimental 
mobility in liquid Ar and CH$_{4},$ but is not very accurate in the 
high--field region \cite{kaneko,kaneko2}.

A great improvement for the description of the mobility in this 
region has been obtained by introducing an energy--dependent 
energy--transfer cross section of the form \cite{BIC}
\begin{equation}
    \sigma_{E}\left( \epsilon \right) = 
    \sigma_{E_{0}} \left({\pi k_{\mathrm{B}}T\over \epsilon} \right)
    \label{eq:eq3}
 \end{equation}
where $k_{\mathrm{B}}$ is the Boltzmann constant and $T$ is the 
temperature.

This particular choice is heuristic and does not rely on any theory. 
However, it is interesting to note that $\pi k_{\mathrm{B}}T/\epsilon$ 
is the square of the ratio of the de Broglie wavelength of an electron of energy 
$\epsilon,$ $\lambda = h/\sqrt{2m\epsilon},$ to its thermal value, 
$\lambda_{T} = h/\sqrt{2m\pi k_{\mathrm{B}}T}.$ In other words, it 
seems that upon collisions the electron is exchanging energy with a 
cross section proportional to the cross sectional area of 
the associated quantum wavepacket. In order to test this view, 
mobility measurements should be carried out as a function of 
temperature. Anyway, we do not insist on this point since it is 
merely speculative.

With the introduction of the analytic form of $\sigma_{E}$ given in Eq. 
\ref{eq:eq3}, the integrals in Eqns. \ref{eq:eq1} and \ref{eq:eq2} can 
be solved analytically, yielding
\begin{equation}
    \mu=\mu_{0}\left[ 1+ \left({M\over 6\pi m}\right) \left({eE\over 
   N k_{\mathrm{B}}T}\right)^{2}\left( 
   \sigma_{m}\sigma_{E_{0}}\right)^{-1}\right]^{-1/2}
    \label{eq:eq4}
\end{equation}
with the usual expression for $\mu_{0}$
\begin{equation}
    \mu_{0}={4e\over 3N\sigma_{m}\sqrt{2\pi mk_{\mathrm{B}}T}}
    \label{eq:eq5}
\end{equation}
It is easy to show that with this choice of $\sigma_{E},$ at 
high--fields $\mu \propto E^{-1},$ so that the drift velocity turns 
out to be approximately constant, as experimentally observed.     

The extension of this model to binary mixtures is easily 
accomplished by introducing the concept of an equivalent one--component fluid with 
density $N$ and mass $M$ of the pure solvent (in this case, Kr) 
but described by effective scattering cross sections $\sigma_{m}(x)$ 
and $\sigma_{E_{0}}(x),$ which now depend on the solute 
concentration \cite{BIC}. 

We have therefore analyzed the present data according to Ref. 
\cite{BIC}, by assuming, for each 
mixtures of concentration $x,$ that the momentum transfer cross 
section is energy independent $\sigma_{m}(x,\epsilon )\equiv 
\sigma_{m}(x),$ and that the energy--transfer cross section is 
proportional to the inverse electron energy as described by Eq. 
\ref{eq:eq3} with $\sigma_{E_{0}}\equiv \sigma_{E_{0}}(x).$

Eq. \ref{eq:eq4} has been fitted to the present experimental data 
with $\sigma_{m}$ and $\sigma_{E_{0}}$ as adjustable parameters. In 
figure 4 we show the values of the parameters resulting from the fit. 

The momentum 
transfer cross section strictly reflects the behavior of $\mu_{0}$ as 
a function of the methane concentration. This is obvious by inspecting 
Eq. \ref{eq:eq5}. Its value is $\sigma_{m}\approx (6\div 7)\times 
10^{-2}$ \AA$^{2}$ and raises up to $\approx 9\times 10^{-2}$ 
\AA$^{2}$ at the largest methane concentration. These values can be 
compared to the value $(0.1 \div 0.4)$ \AA$^{2}$ of the momentum 
transfer scattering cross section of atomic Kr at the Ramsauer 
minimum around $0.5\div 0.7\, eV $ \cite{zecca}.

In the present mixtures, $\sigma_{m}$ is approximately 3-4 
times smaller than 
the value $\sigma_{m}\approx 
0.2$ \AA$^{2}$ obtained for pure liquid Ar \cite{BIC}, as a result of the fact that also in pure liquid Kr 
electrons are more mobile than in pure liquid Ar \cite{yoshino76}.
The behavior of $\sigma_{E_{0}}(x)$ is very different in comparison 
with that of $\sigma_{m}(x).$ It shows a 
large and almost linear increase with increasing  methane concentrations. 
At $x=0,$ 
$\sigma_{E_{0}}\approx 2$ \AA$^{2},$ to be compared to the value 
$\sigma_{E_{0}}\approx 10$ \AA$^{2}$ found in pure liquid Ar 
\cite{BIC}. For $x=10 \,\% ,$ $\sigma_{E_{0}} \approx 170$ \AA$^{2}.$ This 
confirms the fact that the increase of the concentration of methane in 
liquid Kr strongly enhances the inelastic processes leading to 
electron energy relaxation, while it has a nearly negligible effect on 
the processes of momentum transfer.  

It is not completely surprising that the energy transfer cross section 
is much larger than the momentum transfer one because, as already 
pointed out in literature \cite{BIC,lek1,lek2}, the efficiency of 
energy transfer for thermal electrons is larger than that of momentum 
transfer by a factor $\approx1/S(0)\gg 1.$ In any case, we have to 
stress the fact that the behavior of the cross section with the 
concentration confirms the picture that the addition of a molecular 
solute increases the chance of an electron undergoing scattering.

In figure 5 we show the result of this kind of fit for 
the mixture with $x\approx 5\,\% .$ A similar behavior is found in all 
mixtures.
It is easily observed that Eq. \ref{eq:eq4} correctly fits the data 
only up to intermediate field strengths, of the order of several 
hundreds of $V/cm,$ depending on the solute concentration. For larger 
values, $\mu$ decreases less rapidly with increasing $E.$ Its field 
dependence changes from the $E^{-1}$ behavior, predicted by Eq. 
\ref{eq:eq4} with $\sigma_{E}$ given by Eq. \ref{eq:eq3}, to a 
softer $E^{-1/2}$ behavior. This effect is observed for all mixtures.

This new kind of field dependence  at large fields is typical 
for constant cross sections. Therefore, 
at larger fields, hence at larger mean electron energies, the 
scattering events determining the mobility resemble those due to a 
gas of hard spheres. However, we have to recall that this kind of language is 
more pictorial than real, because we are considering effective cross sections 
dressed by many body and multiple scattering effects in a gas--kinetic 
model rather than real two--body collisions.

In the same Figure 5 we show the prediction of the 
gas--kinetic model where we have introduced a constant $\sigma_{E},$ 
in order to simulate the results of hard--sphere scattering. 
The two curves in the figure have been obtained with 
$\sigma_{E}=\sigma_{E_{0}}$ (dashed line), with the value of $\sigma_{E_{0}}$ determined by the 
fit with Eq. \ref{eq:eq4}, and with $\sigma_{E}= (2\pi 
/3)\sigma_{E_{0}}$ (dashed--dotted line), i.e., at thermal energy.  In both 
cases, the calculated mobility at high fields is nearly parallel to the experimental data.

In any case, the deviation of the mobility from the $E^{-1}-$behavior 
towards the $E^{-1/2}-$one indicates that the processes of inelastic 
energy transfer are changing somewhat with the electron energy, as 
though different inelastic channels were opened by increasing the 
mean electron energy, leading to a different energy dependence of 
$\sigma_{E}.$
 
For the sake of completeness, it has to be noted that this change of behavior, 
though less relevant, is observed also in pure liquid Kr. 
A possible explanation of the effect 
in the pure liquid might be related to the possible existence of the 
Ramsauer minimum of the cross section also in the liquid, as argued by 
Christophorou {\it et al.} \cite{christo}. In this case, at high 
fields, the electron energy distribution function is very broad and 
the cross section would be averaged across the Ramsauer minimum, thus 
yielding approximately a constant value.

A careful inspection of the $\mu (E)$ data shows that $\mu$ deviates 
from the $E^{-1}-$behavior when it approximately has the same value 
(within a factor of order unity)
for each mixture. In other words, the deviation takes place when the 
mean electron energy is approximately the same in all mixtures, 
independently of the methane concentration. This confirms the 
hypothesis that new inelastic channels related to the molecular 
impurity open up when the mean electron energy exceeds a given 
threshold.

To give an estimate of the effect, in figure 6 we plot 
the value $E^{\star}$ of the field where $\mu $ takes on the value 
$\mu^{\star}$ and starts deviating from 
the $E^{-1}-$ behavior. 
This threshold electric field has a nice square--root dependence on 
the CH$_{4}$ concentration. This is easy to understand analytically 
if one inverts 
Eq. \ref{eq:eq4} with $\mu^{\star} / \mu_{0}(x) $ and 
$\sigma_{m}(x) $ approximately constant, thus obtaining
\begin{equation}
E^{\star} =\left\{ {6\pi m\over M}
\left[ \left( {\mu_{0}\over \mu^{\star}}\right)^{2}-1 \right]
\left({Nk_{\mathrm{B}}T\over e}\right)^{2}\sigma_{m}\sigma_{E_{0}}
\right\}^{1/2}
\label{eq:eq6}\end{equation}
By inspecting figure 4 one observes that 
$\sigma_{E_{0}}(x)$ increases linearly, to a very good approximation, 
with $x.$ Hence, $E^{\star}$ is approximately proportional to 
$x^{1/2}.$
This observation is another confirmation of the fact that electrons 
are more efficiently thermalized by increasing the impurity content 
of the mixture.

Within this gas--kinetic model, the mean electron energy can be 
calculated as $\langle \epsilon\rangle = 
\int_{0}^{\infty}z^{3/2}g(z)\mathrm{d}z,$ where 
$g(z) $ is given by Eq. \ref{eq:eq2}. If the analytic form Eq. 
\ref{eq:eq3} for the energy--transfer cross section and a constant 
momentum--transfer cross sections are used, the mean 
electron energy turns out to be given by
\begin{equation}
    \langle \epsilon\rangle = {3\over 2} \left[ k_{\mathrm{B}}T 
    +\left( {M\over 6\pi mk_{\mathrm{B}}T \sigma_{m}\sigma_{E_{0}}} 
    \right)\left( {eE\over N} 
    \right)^{2}
    \right]
    \label{eq:enave}
\end{equation}
At the field $E^{\star},$ where the high--field dependence of the mobility 
changes from $E^{-1}$ to $ E^{-1/2},$ the mean electron energy takes on the value $\langle 
\epsilon\rangle^{\star}$ shown in 
Figure 7.
 Beside a small decrease with increasing concentration, $\langle 
\epsilon\rangle^{\star}$ is close to 0.15 $eV,$ quite consistent 
with the value 0.16 $eV$ reported in literature\cite{sakai,sakai2} 
for the excitation of 
the main vibrational level of CH$_{4}.$
 This result further confirms the conclusion
 that the change of the high--field behavior of 
 the mobility is related to the opening of molecular inelastic channels of energy 
 transfer.

\section{Conclusions}\label{sec:concl}
As observed in many other liquid mixtures, also in CH$_{4}-$doped 
liquid Kr the addition of molecular impurities increases 
the efficiency of electron thermalization. This fact permits to 
increase the drift velocity of electrons in the mixtures used as 
sensitive media in ionization detectors with respect to case of the pure liquid.

The use of the gas--kinetic approach also in liquid mixtures to describe the electric field 
dependence of the electron mobility gives very nice results. It is a 
very simple model that relies on a very easy physical picture to 
grasp, namely a binary--collisions picture. 
Nonetheless, in spite of its simplicity, the gas--kinetic model gives 
useful information about the relative strength of elastic and 
inelastic processes through the magnitude and concentration dependence 
of the effective cross sections, $\sigma_{m}$ for the 
momentum--transfer and $\sigma_{E}$ for the energy transfer.

In the case of CH$_{4}-$doped liquid Kr, the molecular solute has a 
relatively small influence on the momentum--transfer processes, as deduced from 
the fact that the 
zero--field mobility $\mu_{0}$ does not depend very much on the 
methane concentration. 
On the contrary, CH$_{4}$ impurities originate an efficient energy 
relaxation of excess electrons. 
This fact has the consequence that collisional ionization of the 
liquid mixture should be more difficult to reach, yielding higher 
breakdown fields in the mixtures than in the pure liquid.
\newpage \section*{Captions to the Figures}
 \begin{itemize}
     
     \item  [Figure 1] Electron mobility in pure liquid Kr (closed 
     circles) and in several liquid Kr--CH$_{4}$ mixtures. Solid 
     triangles: $x=0.5 \,\%.$ Solid squares: $x=5 \,\%.$ Open circles: 
     $x=10 \,\%.$ $x$ is the methane concentration. The error bars, of 
     the order 5--10 \%, are not shown for the sake of clarity.
     
     \item[Figure 2]  Electron mobility in several liquid Kr--CH$_{4}$ 
     mixtures. Closed diamonds:  $x=0.01 \,\%.$ Solid 
     triangles: $x=0.1 \,\%.$ Open squares: $x=1 \,\%.$ Closed 
     squares: 
     $x=2 \,\%.$ Open circles: $x=3 \,\%.$  $x$ is the methane concentration. 
     The error bars, of 
     the order 5--10 \%, are not shown for the sake of clarity.
     \item  [Figure 3] Zero--field electron mobility as a function of 
     the concentration $x$ of CH$_{4}$ in liquid Kr. The solid line is 
     a parabolic fit to the data drawn to guide the eye.
 
     \item  [Figure 4] CH$_{4}$ concentration--dependent behavior of the 
     momentum--transfer cross section $\sigma_{m}$ (closed circle) and of 
     proportionality constant $\sigma_{E_{0}}$ of the energy--transfer cross section
     $\sigma_{E}.$ The solid lines are only guidelines for the eye.

     \item  [Figure 5] Excess electron mobility as a function of the 
     electric field in the mixture with $x=5\, \%$ of methane. Solid 
     line: fit of the kinetic model with the energy--transfer cross section inversely 
     dependent on the electron energy. Dashed-- and dot--dashed lines: 
     prediction of the kinetic model with two different values of an 
     energy--independent energy--transfer 
     cross section.
 
     \item  [Figure 6] Concentration--dependent behavior of the electric 
     field values $E^{\star}$ beyond which the mobility data are no 
     longer well described by an energy--transfer cross section 
     inversely proportional to the electron energy. The solid line is a 
     square--root fit to the data. 

      \item [Figure 7] Dependence on the CH$_{4}$ concentration of the 
      mean electron energy $\langle \epsilon\rangle^{\star}$ evaluated at the field 
      $E^{\star}$ where the high--field behavior of the electron 
      mobility changes from the $E^{-1}$ to the $E^{-1/2}$ dependence. 
      The solid line is only a guide for the eye.
 \end{itemize}


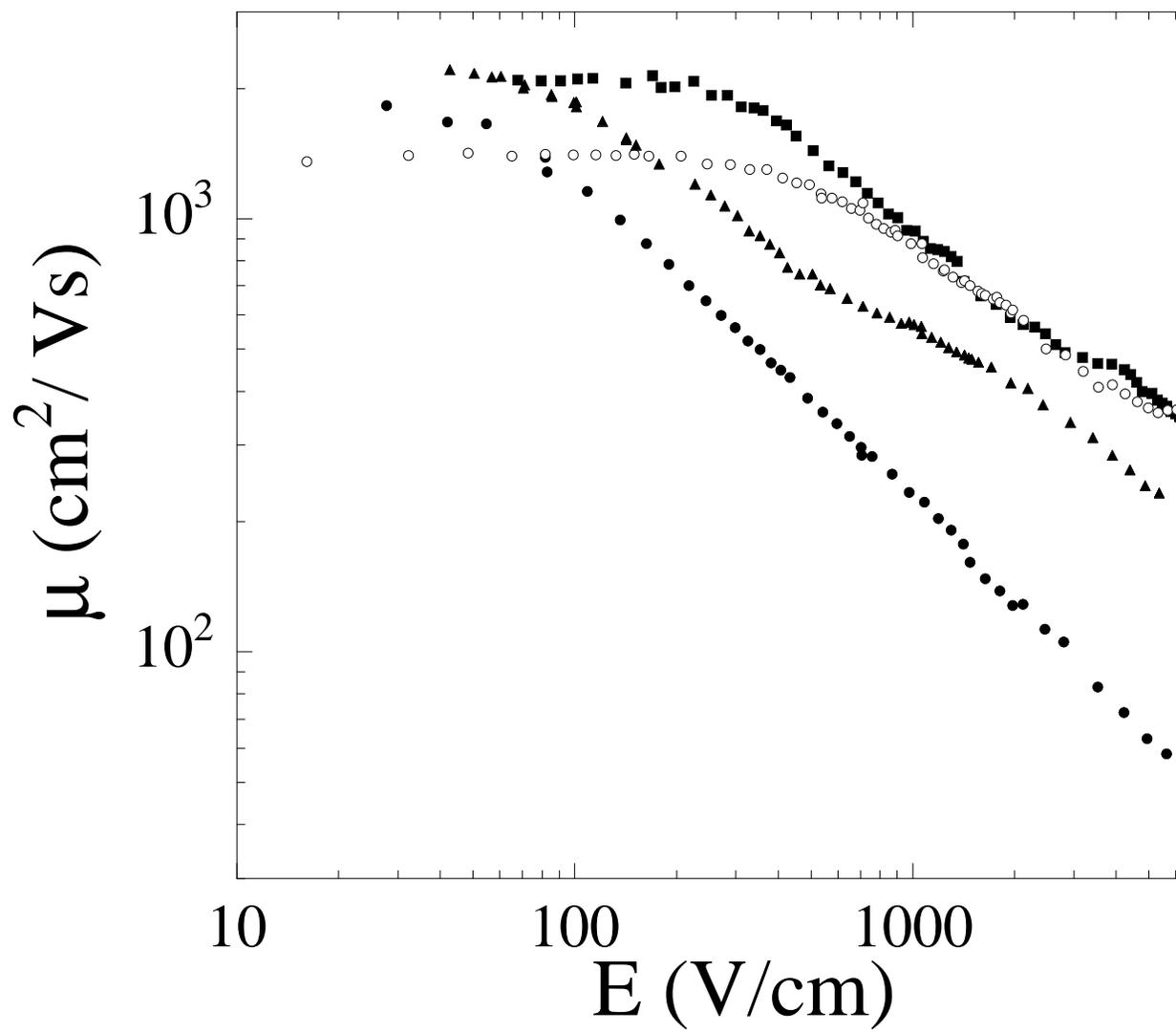
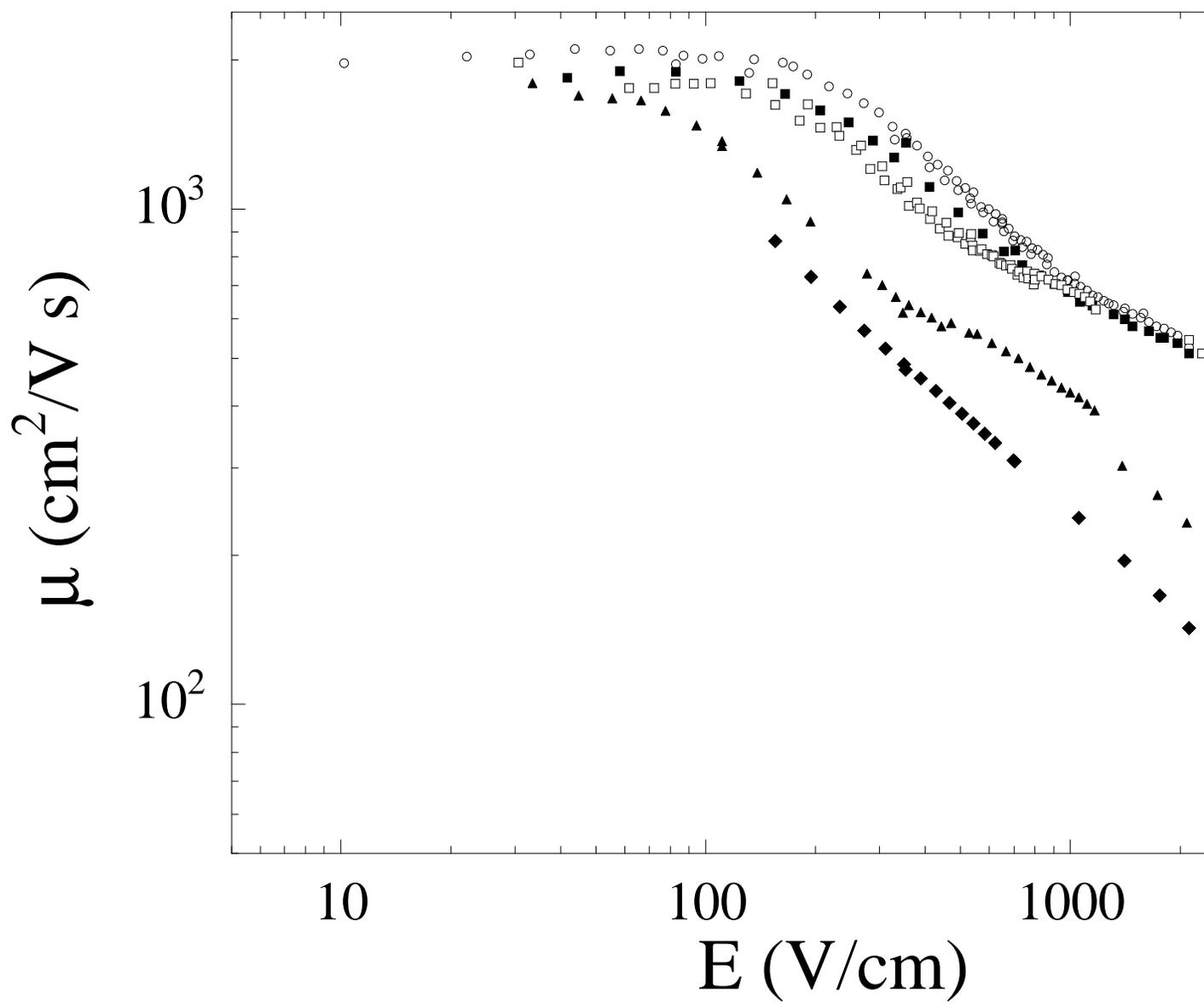
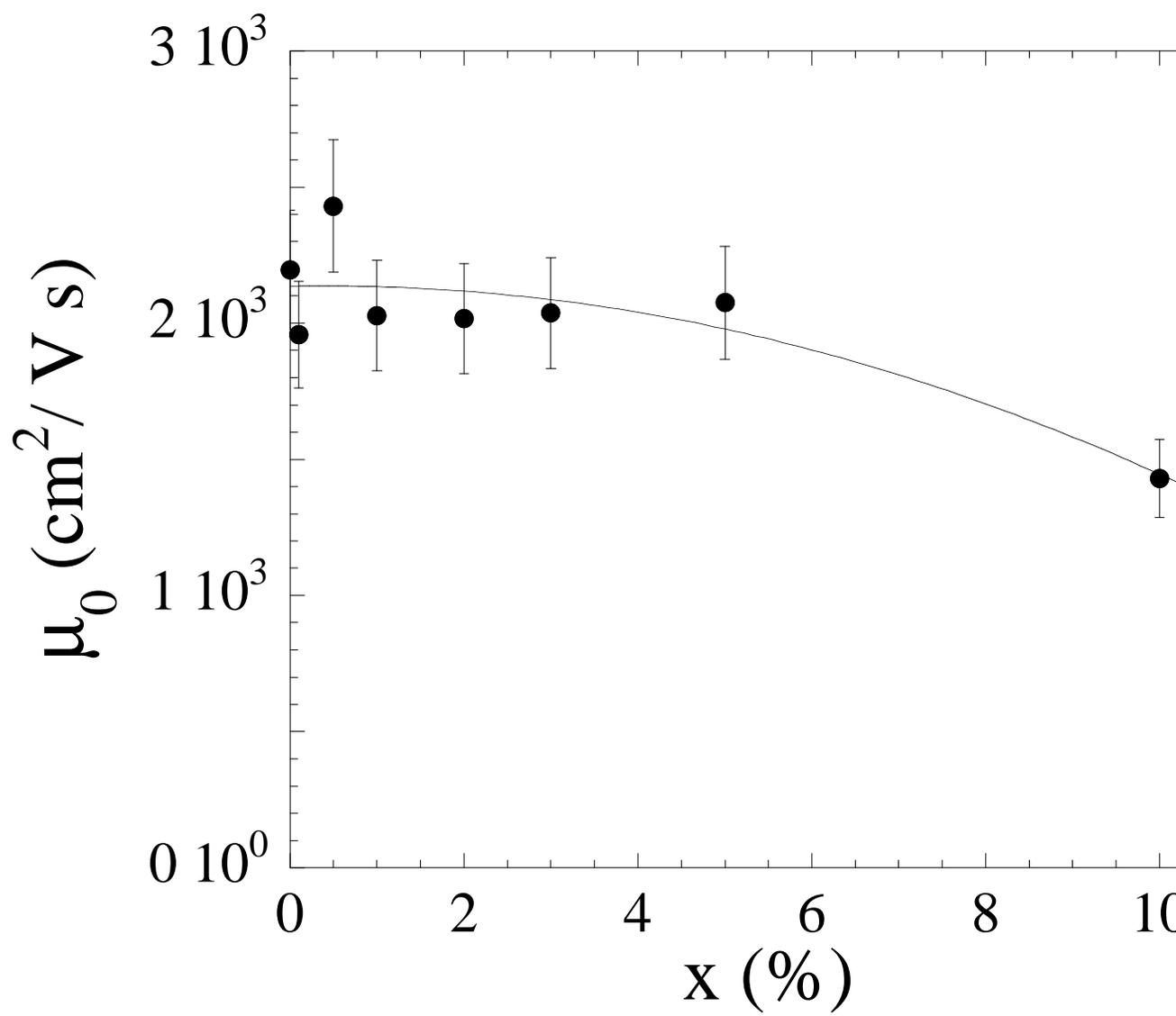
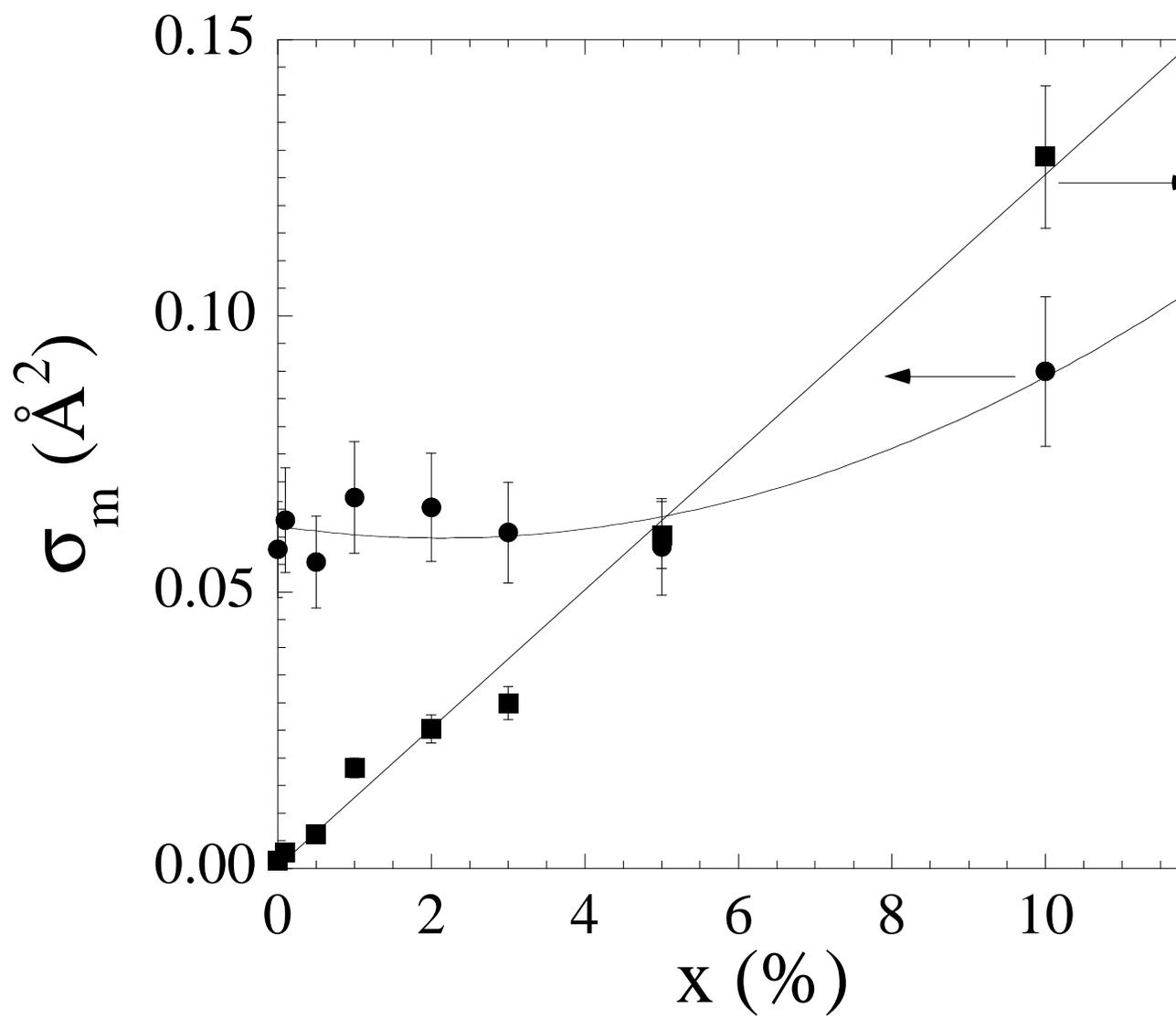
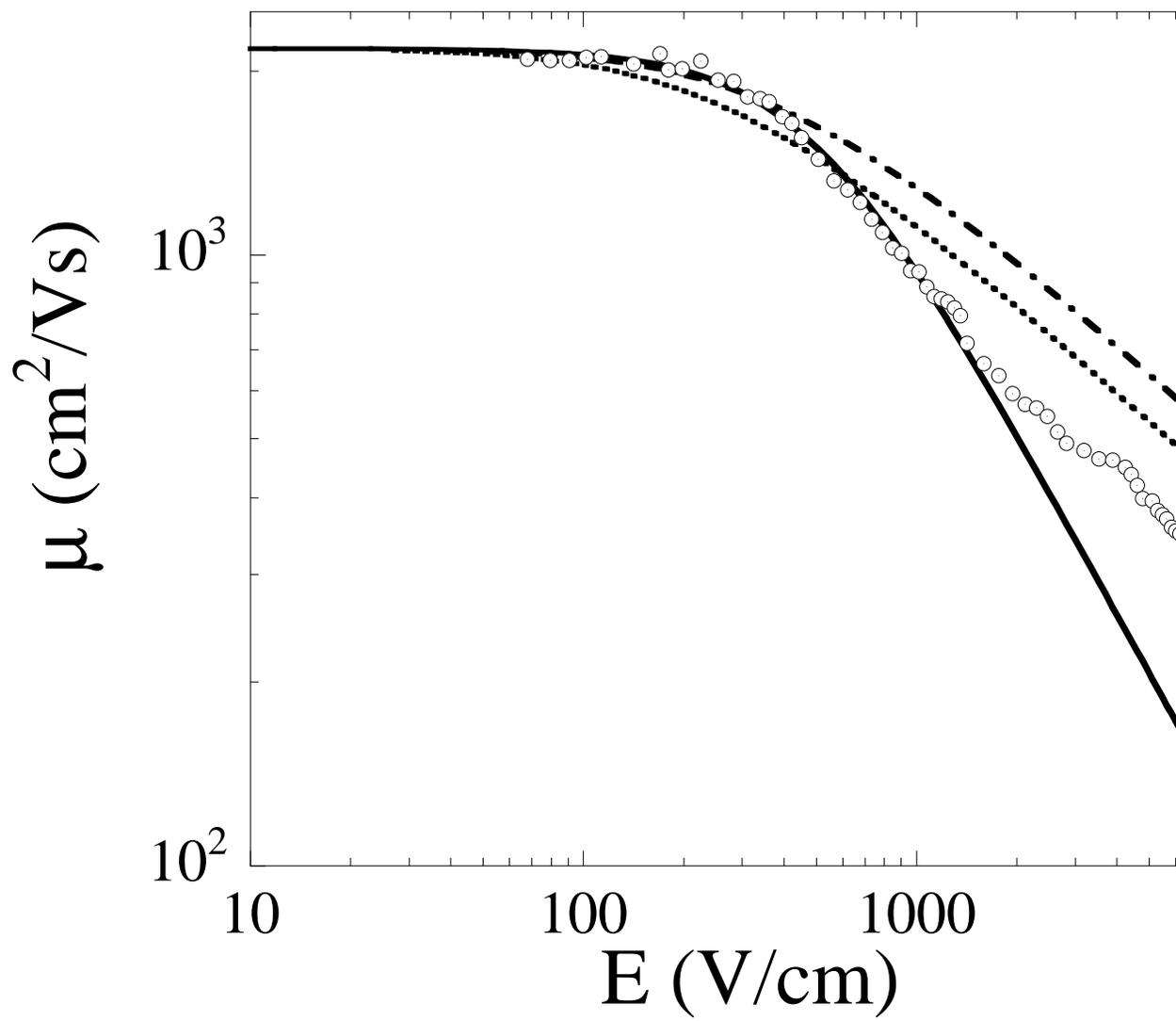
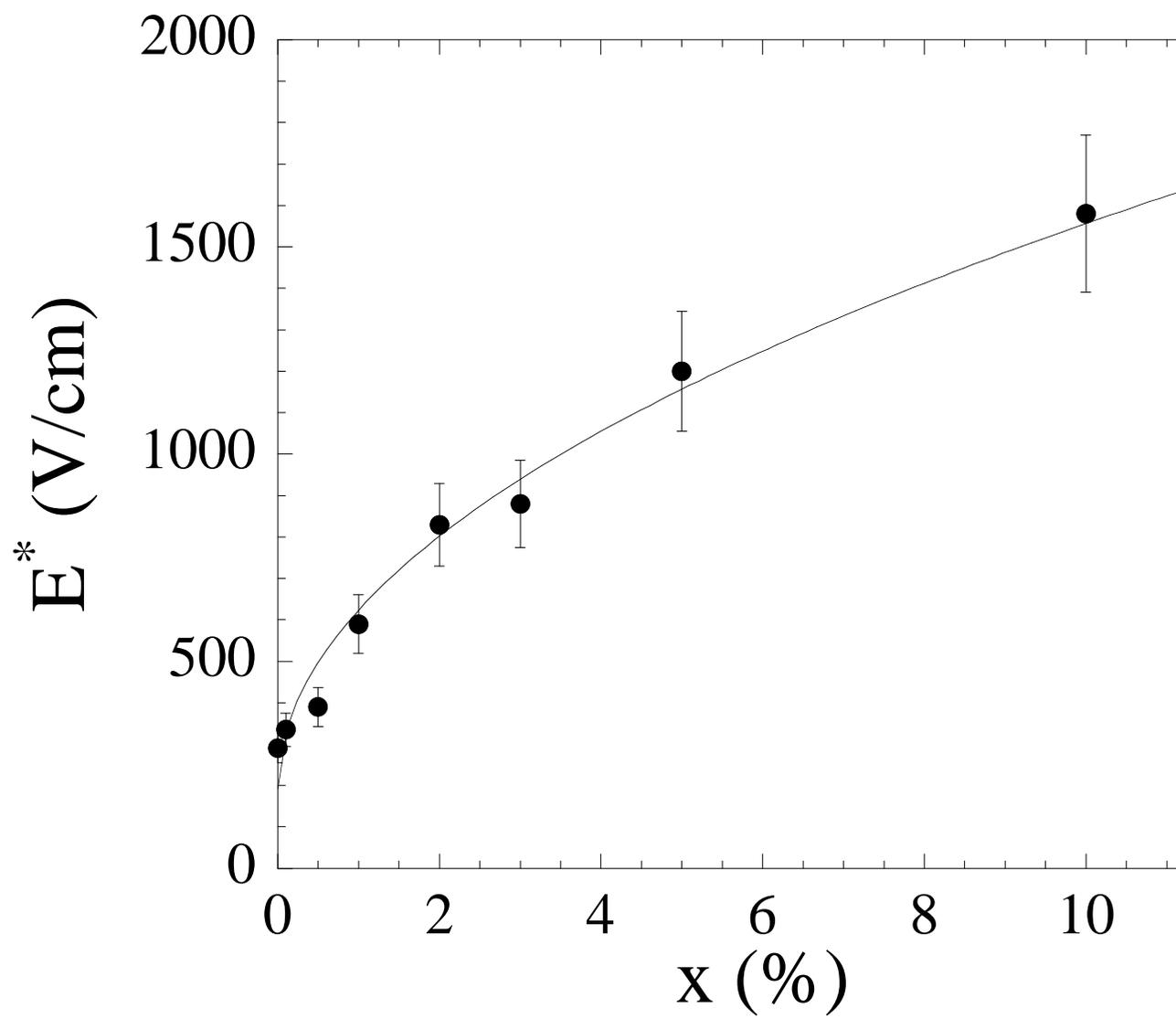
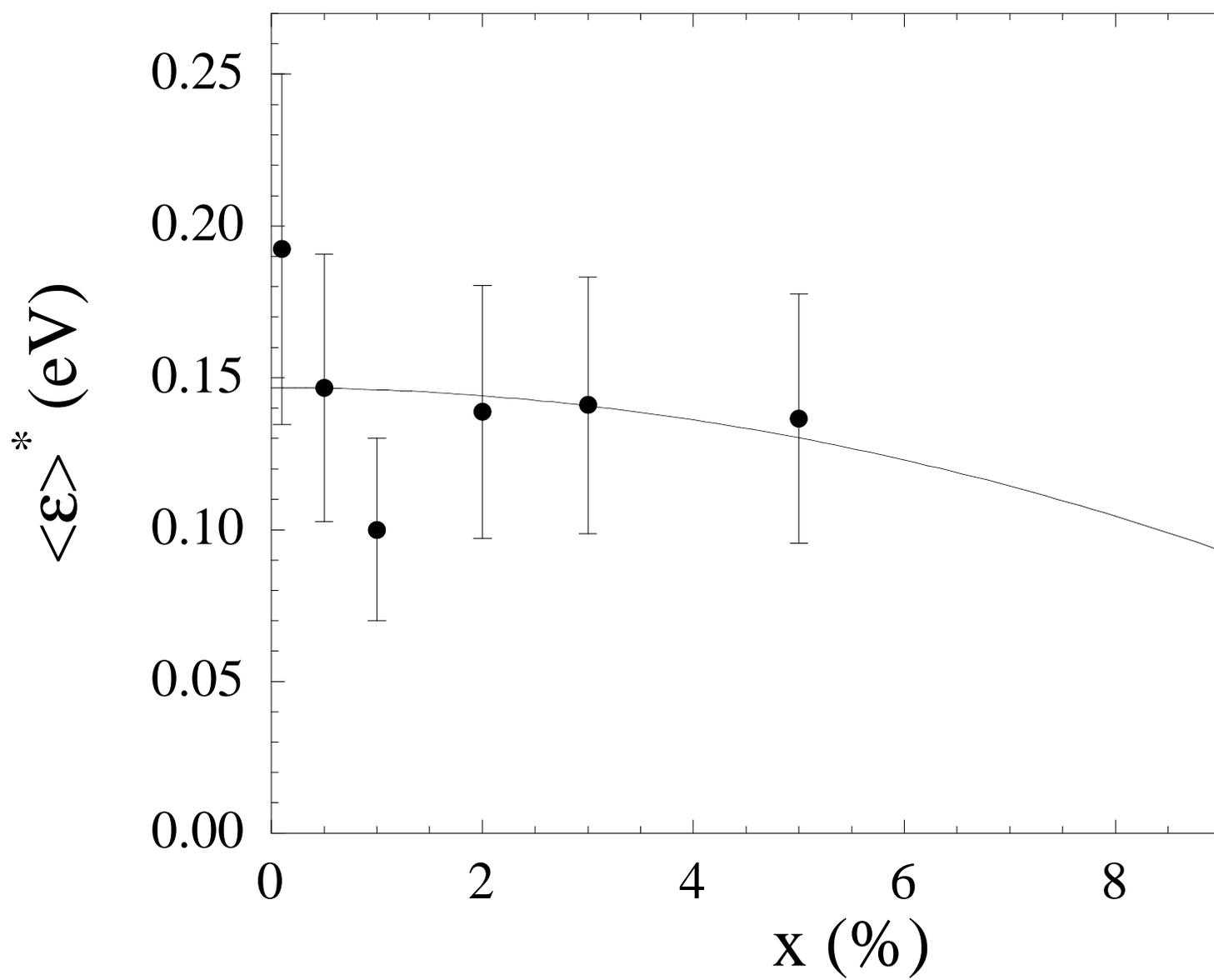
\end{document}